# Toward Generalizable Multiple Sclerosis Lesion Segmentation Models

**LIVIU BADEA**[1,3] and **MARIA POPA**[2,1,3]

[1]Artificial Intelligence Lab, National Institute for Research and Development in Informatics, Romania (e-mail: badea.liviu@gmail.com)
[2]Department of Computer Science, Faculty of Mathematics and Computer Science, Babes Bolyai University Cluj-Napoca, Mihail Kogălniceanu 1, Cluj-Napoca, Romania (e-mail: maria.popa@ubbcluj.ro)
[3]The authors contributed equally to this work

October 25, 2024

## Abstract

Automating Multiple Sclerosis (MS) lesion segmentation would be of great benefit in initial diagnosis as well as monitoring disease progression. Deep learning based segmentation models perform well in many domains, but the state-of-the-art in MS lesion segmentation is still suboptimal. Complementary to previous MS lesion segmentation challenges which focused on optimizing the performance on a single evaluation dataset, this study aims to develop models that *generalize* across diverse evaluation datasets, mirroring real-world clinical scenarios that involve varied scanners, settings, and patient cohorts. To this end, we used all high-quality publicly-available MS lesion segmentation datasets on which we systematically trained a state-of-the-art UNet++ architecture. The resulting models demonstrate consistent performance across the remaining test datasets (are *generalizable*), with *larger* and *more heterogeneous datasets* leading to better models. To the best of our knowledge, this represents the *most comprehensive cross-dataset evaluation of MS lesion segmentation models* to date using publicly available datasets. Additionally, explicitly enhancing dataset size by merging datasets improved model performance. Specifically, a model trained on the combined MSSEG2016-train, ISBI2015, and 3D-MR-MS datasets surpasses the winner of the MICCAI-2016 competition. Moreover, we demonstrate that the generalizability of our models also relies on our original use of *quantile normalization* on MRI intensities.

*Keywords*

Multiple Sclerosis (MS), Brain MRI, Lesion Segmentation, Deep Learning, UNet++, Model Generalizability, Quantile Normalization.[1]

## 1 Introduction

The prevalence of brain disorders has surged in the past century, becoming increasingly widespread across all age groups. Conditions such as stroke, epilepsy, Alzheimer's and Parkinson's disease, glioblastoma, and multiple sclerosis have become notably common. Multiple Sclerosis (MS) is an autoimmune demyelinating disease impacting both the brain and spinal cord, typically affecting young adults and more frequently females. According to the World Health Organization (WHO), approximately 1.8 million individuals worldwide have been diagnosed with multiple sclerosis.[2] This disorder affects various cognitive, emotional, motor, sensory and visual functions. Magnetic Resonance Imaging (MRI) stands as the sole objective screening method capable of detecting MS. Regular followups must be conducted to monitor the progression of the disease and to determine a treatment plan aimed at slowing its advancement. Medical professionals are required for identifying demyelinating lesions, quantifying them, and tracking disease progression through longitudinal screenings. Due to the laborious nature of the task of *quantitatively* tracking the lesions by expert diagnosticians, there is an obvious necessity for a computer-assisted system to streamline the workload of medical staff, by devising an accurate lesion segmentation method.

---

[1]This work has been submitted to the IEEE for possible publication. Copyright may be transferred without notice, after which this version may no longer be accessible.

[2]`www.who.int/news-room/fact-sheets/detail/multiple-sclerosis`

Traditional image analysis and computer vision methods, such as histogram-based, graph-based, and thresholding methods, are prone to failure in multiple sclerosis segmentation tasks. This is due to the fact that demyelinating lesions are not solely characterized by hyper-intensity, but also by the surrounding image context, which is challenging to define a priori using specific predefined features.

The emergence of *deep learning methods* [1] has transformed the realm of computer vision by their ability to automatically infer appropriate features and representations that can directly aid in solving the problem. However, this requires vast quantities of data, which unfortunately existing MS datasets do not provide. For instance, while datasets like ImageNet [3] boast millions of images featuring diverse object classes, available MS data consist of only thousands of brain slices, with substructures that are much harder to differentiate than those from real life images.

The MS lesion segmentation literature contains various deep learning approaches that operate on diverse datasets [2, 3, 4, 5, 6, 7]. Despite several competitions, which encouraged model optimization on a fixed test dataset, the problem persists due to inadequate numbers of samples as well as the variations in MRI hardware and parameters. Notably, several such competitions were organised by MICCAI in 2008 and 2016 [8, 7], and by the International Symposium on Biomedical Imaging (ISBI) in 2015 [9]. In 2021, MICCAI once again spearheaded a multiple sclerosis challenge, with a particular emphasis on longitudinal screenings [10]. The rather suboptimal results of these competitions suggest there is a noticeable dearth of training data that cannot be properly compensated by even the most sophisticated algorithms (the latter become effective only when sufficient data becomes available). Various approaches were tested, resulting in approximately similar outcomes. Other approaches trained diverse architectures on in-house data and subsequently fine-tuned the models to fit the dataset utilized in the competition, but the results are not nearly good enough to be of clinical use (see the leaderboard for the MICCAI 2016 competition [7]).

As previously noted, there are few publicly available datasets for MS lesion segmentation and most are obtained in uniform setups. MSSEG2016 is the only heterogeneous dataset, comprising data from multiple centers and MRI machines, albeit with a limited number of samples.

To our knowledge, only a few studies have trained and tested models on more than one or two datasets. Their primary objective has been to *optimize results on a single test dataset*, which is a valid approach given that competitions typically emphasize model performance. However, optimizing on a single dataset may impact the model's generalizability across multiple datasets obtained in heterogeneous setups. Moreover, they have not employed a cross-dataset approach for training and testing, which would involve training on one dataset (e.g., A) and testing on the remaining datasets (e.g., B, C, D, ...), and repeating this process for all permutations. Without a comprehensive analysis of *how the models perform across multiple datasets*, the direct clinical applicability remains uncertain. This is because there is no guarantee that these models can be effectively utilized in clinical settings, that may be distinct from the settings used to obtain the training data.

This paper focuses on investigating the *generalizability of models* through systematic training and testing on multiple datasets. Note that generalizability is not a straightforward consequence of a highly performing model optimized on some training dataset, as can be seen for example in [11] (see also Section 5.1). To address this, we gathered as many MS datasets as possible, studying the abilities of models to generalize across various datasets rather than their performance on a single test dataset. This approach provides more confidence in the robustness of such models in distinct clinical settings.

We show that the learned models behave well broadly across a larger number of datasets. We also show that by combining multiple datasets we obtain models that match and even surpass the winner of the MICCAI 2016 competition without any hyperparameter optimizations. Moreover, we demonstrate that the generalizability of our models is influenced by the normalization of MRI intensities. We employ *quantile normalization*, a method initially designed for normalizing genomic data, which, to the best of our knowledge, is being applied to MRI data for the first time. We also investigate the influence of expert annotation quality on model performance, given that the agreement between expert annotations is far from perfect (around 60-70%).

The key contributions of the paper are as follows:

- We present a comprehensive analysis of MS lesion segmentation using all high-quality publicly available datasets with the focus on building generalizable models. Compared to previous work, we push the study of generalizability one step further, by showing that models trained on *any* of the existing public datasets generalize well to the others.

---

[3]ImageNet: www.image-net.org

- We show that quantile normalization of heterogeneous intensity distributions (of scans obtained with different scanners and setups) improves model generalizability.

- We show that a simple architecture trained on limited data of a single MRI modality can surpass the official winner of the MICCAI 2016 competition, while generalizing well to other datasets. Whereas previous lesion detection algorithms have demonstrated strong performance using internal datasets that are an order of magnitude larger and incorporating several MRI modalities (e.g., FLAIR and T1w), we demonstrate that similarly robust results can be achieved with a simpler method and just a single MRI modality (FLAIR) even with an order of magnitude less training data.

The rest of the paper is structured as follows: Section 2 discusses the related work, Section 3 provides an overview of the publicly available datasets, Section 4 introduces the methodology, while Section 5 presents the findings, followed by concluding remarks and future directions.

## 2 Related work

MS lesion segmentation research goes back more than a decade, with a major MICCAI challenge organised already in 2008 [4]. For brevity, we concentrate in the following on the recent state-of-the-art.

The study by Rondinella [3] introduced a fully convolutional DenseNet with attention, which was trained on the ISBI2015 challenge dataset [9]. A single scan of a patient (out of several scans acquired for this patient at different timepoints) was left out for testing purposes. Since the test data is not completely independent from the training data (scans from the same patient, albeit at different timepoints, occur both in the training and the test data), significant data leakage is possible due to the potentially significant overlaps of the lesions at different timepoints. In a similar vein, [2] uses a modified UNet architecture and an essentially identical testing method as [3]. Both approaches reported Dice scores exceeding 0.8 on the test patient (0.89 [3] and respectively 0.82 [2]).

Sadeghibakhi et al. [5] applied the Contrast Limited Adaptive Histogram Equalization (CLAHE) algorithm to process the MRI images, followed by the utilization of a Laplacian detector to extract edges. The original image and the one obtained from the edge detector were concatenated into a 4D structure, whose patches with dimensions 80x80x80x2 were then employed for training a modified attention UNet architecture. Similarly to [3] and [2], the ISBI2015 dataset was utilized for training. However, for testing, a set of unseen data proposed by the competition was employed. Using independent test data yielded a Dice score of 0.64, which is likely a better estimation of the true performance of the model.

The report on the MICCAI 2016 multiple sclerosis challenge [7] details the participation of 13 teams. Each team pursued a distinct approach, employing methods such as graph cuts, neural networks, ensemble techniques and random forests. Some teams used models pretrained on in-house datasets. However, while the methods demonstrated promising results on the designated test dataset, there remains a gap in evaluating the robustness of these methods on other independent datasets. The top four teams obtained Dice scores ranging from 0.52 to 0.59.

In a recent study, Sarica et al. [6] introduced a dense residual UNet for multiple sclerosis lesion segmentation. Unlike previous literature, this study conducted more comprehensive training and testing scenarios involving two datasets, ISBI2015 and MSSEG2016. Initially, a model was trained on the ISBI2015 dataset and tested on its corresponding test data, resulting in a Dice score of 0.6688. Subsequently, a model underwent training on the MSSEG2016 dataset and was tested on the MSSEG2016-test dataset achieving a Dice score of 0.6727. Furthermore, the study also presents a cross-dataset validation scenario where the method was trained on ISBI2015 and tested on MSSEG2016, and vice versa, yielding lower Dice scores, of 0.6031 and 0.4819, respectively.

Another notable work [11] introduced DeepLesionBrain (DLB), a novel method for multiple sclerosis lesion segmentation that is robust to domain shifts, as demonstrated through cross-dataset testing. DLB employs a set of compact 3D convolutional neural networks distributed across the brain, each focusing on specific regions with overlapping receptive fields to simplify the complex task of whole-brain segmentation into manageable subtasks, ensuring consistent results. The method incorporates Hierarchical Specialization Learning (HSL) to extract both global and local features by first training a generic network and then refining it for specific brain regions. Additionally, DLB utilizes Image Quality Data Augmentation (IQDA) to introduce realistic variations in training data, enhancing the model's robustness to different data sources and quality levels, thus improving generalization across unseen domains. DLB has achieved robust results across a high quality internal and 2 public datasets (ISBI2015, MSSEG2016) using two MRI modalities (T1w and FLAIR). Our work shows that similarly robust results can be obtained using a single MRI modality (FLAIR) using a much simpler method.

---

[4] www.nitrc.org/projects/msseg

Recently, [12] released a model, LST-AI, trained on an internal dataset of 491 pairs [5] of FLAIR and T1w scans. LST-AI obtained Dice scores ranging between 0.61 and 0.74 on the publicly available datasets (with an average of 0.67), outperforming all previous methods. On the other hand, we are trying to obtain generalizable models using a single imaging modality (FLAIR)[6] and using publicly available datasets, which are much smaller. Moreover, while LST-AI is a single model (actually an ensemble of three models) that generalizes well on the few existing public datasets, we push the study of generalizability one step further, by showing that training a model on *any* of the existing public datasets generalizes well to the others, even though the training datasets are an order of magnitude smaller. In general, generalizability of a model cannot be proven on a limited number n of test datasets, as the model can always fail on another, (n+1)-th dataset. By training models on multiple datasets and demonstrating their robust performance across all test datasets, we increase the confidence in the generalizability of the models and the corresponding training method. For example, [11] showed that training the nicMSlesion algorithm on their largest, in-house dataset obtains a catastrophic Dice score of 0.13 on ISBI2015, despite the high Dice score of 0.65 on MSSEG2016, showing that this particular model is not generalizable (Table 6 of [11]).

In conclusion, the state-of-the-art in MS lesion segmentation is around 0.67 in terms of Dice score for public datasets, with significantly higher scores obtained only in studies where the test data was not independent of the training data or used larger or higher quality internal datasets.

## 3 Datasets

To assess the effectiveness and robustness of the proposed method, the following publicly available datasets were used (see Table 1).

Table 1: Datasets

| Dataset | Scanner | Manufacturer | Subjects | Scans | Slices |
|---|---|---|---|---|---|
| ISBI2015 | 3T | Philips Medical Systems | 5 train | 21 | 2843 |
| MSSEG2016-train | 1.5T & 3T | Center<br>• 01: Siemens Verio 3T<br>• 07: Siemens Aera 1.5T<br>• 08: Philips Ingenia 3T | 15 | 15 | 3021 |
| MSSEG2016-test | 1.5T & 3T | Center<br>• 01: Siemens Verio 3T<br>• 03: General Electrics Discovery 3T<br>• 07: Siemens Aera 1.5T<br>• 08: Philips Ingenia 3T | 38 | 38 | 8289 |
| 3D-MR-MS | 3T | Siemens Magnetom Trio | 30 | 30 | 10517 |

### 3.1 ISBI2015

The training dataset [9] for the ISBI Longitudinal Multiple Sclerosis challenge was made publicly accessible in 2015.[7]. It includes T1w, T2w, PDw and T2-weighted FLAIR images acquired using a 3T MRI scanner. The dataset is longitudinal, featuring 4 to 6 follow-up scans for each patient, spanning an acquisition period of one year. It is divided into training and testing data segments. The training dataset comprises 5 patients with 4 to 5 follow-up MRI scans per patient, totaling 21 MRI scans. Conversely, the testing segment includes 14 patients with 4 to 6 follow-up scans per patient, resulting in a total of 61 scans. Each MRI scan was segmented by two raters, with segmentation masks disclosed only for the training set. Therefore, in this paper only the ISBI2015 training dataset was used (referred to as ISBI2015).

---

[5]At least an order of magnitude larger than all the publicly available datasets.

[6]Paired T1w and FLAIR images may not be available in all clinical settings.

[7]ISBI Challenge website: https://smart-stats-tools.org/lesion-challenge-2015

## 3.2 MSSEG2016

The MICCAI 2016 Multiple Sclerosis lesion segmentation challenge [8] introduced a dataset comprising lesions manually segmented by 7 experts, referred to as MSSEG2016 in the rest of this study. This dataset includes MRIs from 53 patients, 15 of each used for training and the remaining 38 for testing. The data was acquired in 4 centers, each utilizing a distinct scanner. The training set comprises patients from 3 out of 4 centers, while the test set includes patients from all 4 centers, ensuring that one center (center03) remains unseen during training. Both training and test data include lesion segmentation masks from the 7 experts and a consensus mask used as ground truth. Each patient's imaging modalities encompassed a 3D FLAIR image, a 3D T1-weighted image before and after contrast substance (Gadolinium) injection, and an axial dual PD-T2 weighted image, all featuring similar image resolutions. The dataset is accessible upon request on the Shanoir website [8]. As discussed in the official competition report [7], one specific case from the test dataset (patient 8 from center 7) had an empty consensus segmentation, meaning that the 7 experts did not agree on the position and extent of any lesion in that case. Therefore, this case needed to be excluded since no ground truth was available for it.

## 3.3 3D-MR-MS

In [13], a multiple sclerosis database comprising data from 30 patients was released. These data were obtained using a 3T MR scanner with conventional sequences at the University Medical Center Ljubljana (UMCL). For each scan, three expert raters conducted segmentations of white matter lesions, which were subsequently refined through collaborative sessions to establish a consensus segmentation. The scans of each patient include a T1w, a T2w and a FLAIR image. Notably, there is no division between training and testing data. The data can be accessed on the laboratory website [9].

# 4 Methodology

To test the *generalizability* of lesion segmentation models, we comprehensively trained a model on each of the 4 datasets and tested it on the remaining ones, checking if model performance is similar across all training-test dataset combinations (Figure 1).

Since the existing datasets are rather small, we also tested whether explicitly increasing the dataset size by combining datasets improves model performance .

The state-of-the-art model performances are around 0.67. This could be due to the shortage of MS data, but also to the reduced quality of the annotations. Surprisingly, we observed that expert annotators show less-than-perfect agreement in their lesion annotations, prompting us to investigate the extent to which this impacts model performance.

## 4.1 Preprocessing

We used the original preprocessed versions of the datasets, as all underwent similar preprocessing steps, namely skull stripping and N4 bias field correction.

In addition to the standard preprocessing steps, we applied *quantile normalization* to reduce the rather large intensity distribution differences between images acquired with different scanners. Quantile normalization, a method initially designed for genomic data [14, 15], adjusts the distribution of the intensities in an image to the standard distribution of a given template.

Finally, 2D slices (of constant z) in the standard axial orientation were extracted from the 3D images and resized to 224x224. Only slices containing brain tissue were included in the dataset (see Table 1 for the corresponding numbers of slices). Both quantile normalization and the use of standard axial slices aim to mitigate dataset variability and enhance model robustness. For simplicity, we only use the FLAIR images from the datasets, thereby broadening the applicability of our approach. This choice was also motivated by previous studies which have not observed major benefits of using additional MRI modalities [5].

---

[8]MSSEG2016: `https://shanoir.irisa.fr/shanoir-ng/challenge-request`

[9]3D-MR-MS: `https://lit.fe.uni-lj.si/en/research/resources/3D-MR-MS/`

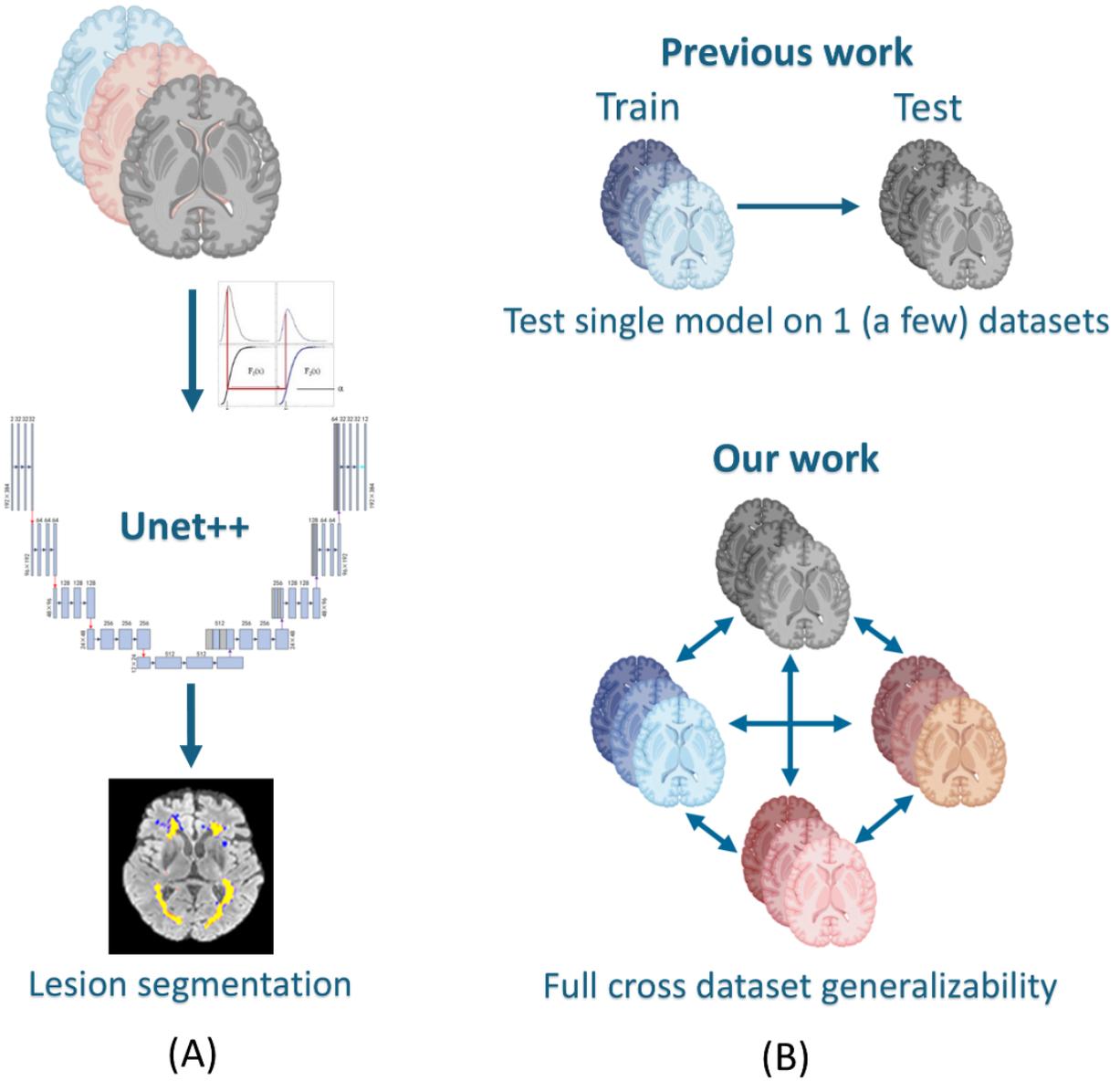

Figure 1: Our approach: (A) MS lesion segmentation architecture. (B) Full cross-dataset testing of generalizability.

### 4.2 Model architecture and training

The state-of-the-art deep learning architectures used in medical image segmentation are UNets [16]. These network architectures employ an encoder-decoder structure, comprising contracting paths for feature extraction and expansive paths for precise localization.

In the UNet++ architecture [17] the encoder and decoder are connected through a series of nested, dense skip pathways instead of the standard UNet skip connections. These skip pathways use dense convolution blocks with the aim of reducing the semantic gap between the feature maps of the encoder and those of the decoder prior to fusion. We chose the UNet++ architecture for its slightly better performance compared to the standard UNet across the chosen datasets. We used the UNet++ implementation from the Segmentation Models PyTorch library [18] and tested several Timm encoders [19], including `resnet18`, `timm-efficientnet-b0`, `eca_nfnet_l0`, and `timm-mobilenetv3_small _minimal_100`. Since `resnet18` yielded slightly better results, it was used in all subsequent runs. 80% of the patient-data was used for training and 20% for model selection (validation). To avoid data leakage, the split was performed such that all slices of a scan are either in the training set or in the validation set.

The chosen model was trained to minimize a weighted BCE Loss, with the Adam optimizer and the following hyperparameters: initial learning rate 1e-3, weight decay 1e-6, betas (0.9, 0.999) and BCE Loss weight 0.8 (to deal with class imbalance). Given the large numbers of parameters of such models and the small dataset sizes, regularization methods (such as weight decay) are essential to avoid model over-fitting. All models were trained for 50 epochs and the best model with the highest Dice score over the validation set was selected. Testing was always performed on a completely independent dataset.

### 4.3 Evaluation metrics

For evaluation purposes, both the Dice coefficient (DSC) and the Jaccard index (IoU) were computed. These metrics quantify the relative overlap between the predicted and the ground truth lesion segmentations of the MRI image:

$$DSC = \frac{2TP}{2TP + FP + FN} \qquad\qquad IoU = \frac{TP}{TP + FP + FN}$$

where

- TP represents the True Positives (the pixels accurately identified as multiple sclerosis lesions)
- TN - True Negatives (the pixels accurately identified as non-lesions)
- FP - False Positives (the pixels erroneously predicted as MS lesions)
- FN - False Negatives (the pixels inaccurately predicted as non-lesions).

Since the DSC is the primary metric reported in the MS lesion segmentation literature and competitions, we mainly compare our models in terms of DSC.

## 5 Results

To assess the generalizability of the selected model across the available datasets, extensive training and testing were conducted.

### 5.1 Lesion segmentation models are generalizable to multiple datasets

First, a model was independently trained on each dataset and subsequently tested on all the others, with the results presented in Table 2. It can be seen that the models yield similar outcomes across all dataset combinations, thereby demonstrating generalizability. To the best of our knowledge this is the most extensive cross dataset evaluation of MS lesion segmentation models using publicly available datasets. Figure 2 shows the lesion segmentations produced by all our models on a given 2D slice of a certain patient scan.

Although extensively studied on single datasets, cross-dataset validation of MS lesion segmentation has been addressed in far fewer publications [6, 11]. While [6] reports the cross-dataset validation of only a single algorithm, [11] performs a more extensive cross-validation of 4 different algorithms (nicMSlesion, DeepMedic, 2.5D Tiramisu and DLB), but only 2 public datasets (MSSEG2016, ISBI2015) and one in-house dataset are considered. The results reported in

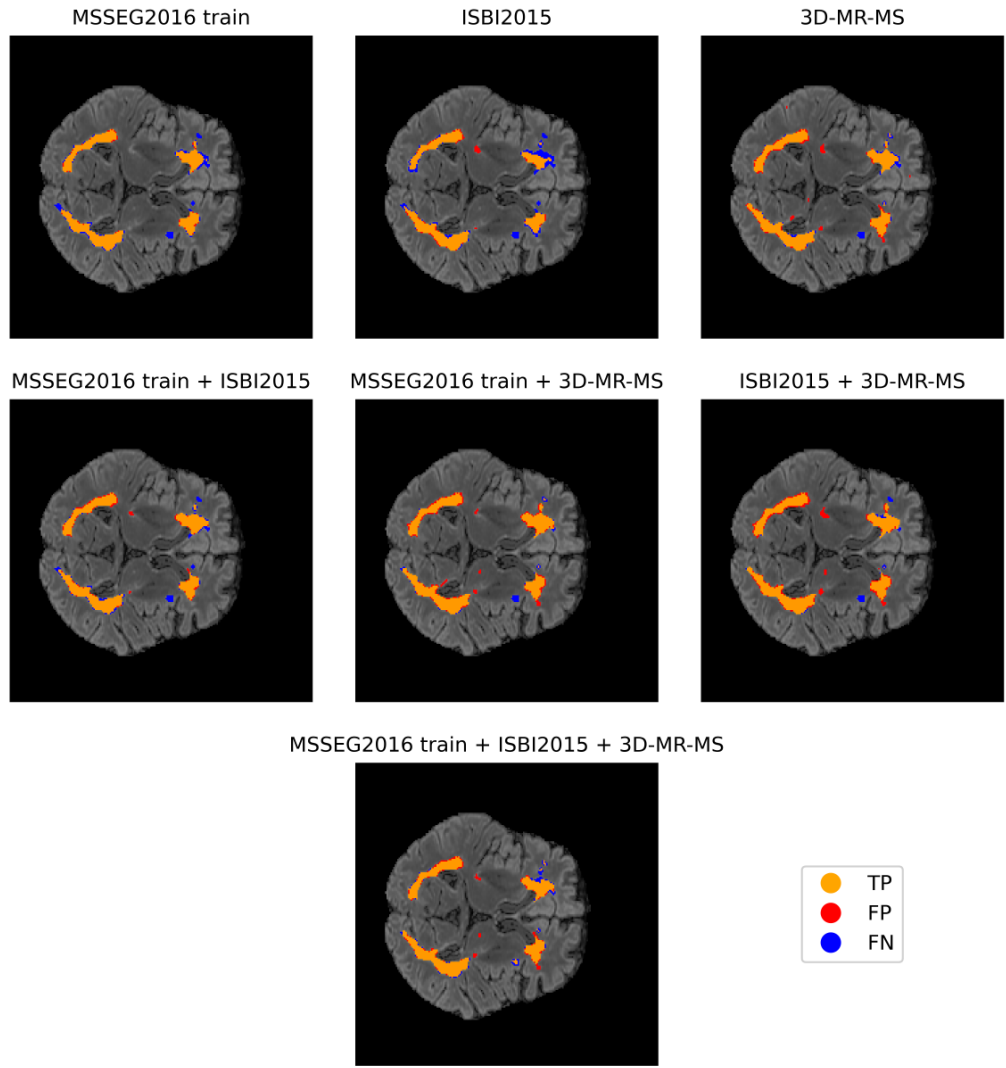

Figure 2: Lesion segmentations produced by our models (trained on different datasets) on slice 148 of the scan of patient 1 from center 3 from the MSSEG2016 test dataset. Orange: True Positives, Red: False Positives, Blue: False Negatives (Ground Truth: Orange+Blue, Predicted: Orange+Red).

Table 2: Performance of models trained on individual datasets and tested on the rest.

| Trained on | Tested on | Dice score | IoU score |
|---|---|---|---|
| MSSEG-2016 train | MSSEG-2016 test | 0.602 | 0.456 |
| MSSEG-2016 train | 3D-MR-MS | 0.523 | 0.379 |
| MSSEG-2016 train | ISBI-2015 | 0.569 | 0.407 |
| 3D-MR-MS | MSSEG-2016 test | 0.544 | 0.406 |
| 3D-MR-MS | MSSEG-2016 train | 0.606 | 0.456 |
| 3D-MR-MS | ISBI-2015 | 0.603 | 0.444 |
| MSSEG-2016 test | MSSEG-2016 train | 0.698 | 0.542 |
| MSSEG-2016 test | 3D-MR-MS | 0.569 | 0.423 |
| MSSEG-2016 test | ISBI-2015 | 0.623 | 0.462 |
| ISBI-2015 | MSSEG-2016 train | 0.515 | 0.352 |
| ISBI-2015 | MSSEG-2016 test | 0.516 | 0.362 |
| ISBI-2015 | 3D-MR-MS | 0.460 | 0.273 |

[11] show that generalizability is not a trivial consequence of a highly performing model optimized on some training dataset. For example, training the nicMSlesion algorithm on the largest (in-house) dataset obtains a catastrophic Dice score of 0.13 on ISBI2015, despite the high Dice score of 0.65 on MSSEG2016, showing that this particular model is not generalizable (Table 6 of [11]). Similar non-generalizable models were obtained by 2.5D Tiramisu trained on MSSEG2016 (with a lower Dice Score of 0.165 on ISBI2015, compared to 0.706 on the in-house dataset) and by nicMSlesion trained on ISBI2015 (with a 0.204 Dice score on the in-house dataset contrasting a 0.442 Dice score on MSSEG2016). Therefore our generalizability results are non-trivial, as a highly performing model on a limited dataset is not guaranteed to perform well on other independent datasets.

## 5.2 Larger and more heterogeneous datasets produce better models

The highest cross dataset Dice score of 0.698 is achieved by our model trained on the 'MSSEG2016 test' dataset and tested on the 'MSSEG2016 train' dataset. This is likely due to the 'MSSEG2016 test' dataset having more samples than 'MSSEG2016 train'. Moreover, although both datasets are heterogeneous (contain scans acquired with different scanners in different centers), the scanner setups from 'MSSEG2016 train' are among those used in 'MSSEG2016 test' (unlike other test datasets considered in Table 2, which are completely independent). The converse is not true, since MSSEG2016-test includes a center (center03) that does not occur in MSSEG2016.

Next, we compared the models trained on different datasets to determine if any characteristics of the training dataset influence their (average) performance on other datasets. To this end, we performed a one-way ANOVA test on the means of the Dice scores of the models from Table 2 (see Table 3 for the corresponding means). The null hypothesis of no difference between means was rejected at p=0.038.[10]

Table 3: Mean Dice scores for models trained on individual datasets and tested on the remaining ones.

| Trained on | Slices | Heterogeneous? | Mean Dice score |
|---|---|---|---|
| MSSEG-2016 train | 3021 | heterogeneous | 0.564 |
| 3D-MR-MS | 10517 | homegeneous | 0.584 |
| MSSEG-2016 test | 8289 | heterogeneous | 0.630 |
| ISBI-2015 | 2843 | homegeneous | 0.497 |

Indeed, *heterogeneous datasets tend to perform better than homogeneous ones* for approximately equal numbers of samples. For example, the heterogeneous 'MSSEG2016 train' dataset (with 3021 slices) performs better (0.564 > 0.497 Dice score) than the homogeneous ISBI2015 dataset (with a comparable number of 2843 slices). Likewise, the heterogeneous 'MSSEG2016 test' dataset (8289 slices) marginally outperforms the homogeneous 3D-MR-MS dataset (10517 slices): 0.630 > 0.584 average Dice score.

---

[10]The post-hoc Tukey HSD analysis revealed a statistically significant difference between the large heterogeneous dataset MSSEG-2016 test and the small homogeneous dataset ISBI-2015, as expected (p=0.02714).

On the other hand, *the models trained on larger datasets tend to outperform those trained on smaller datasets*, as long as both are either homogeneous or heterogeneous. In the case of heterogeneous datasets, 'MSSEG2016 test' (8289 slices) marginally outperforms 'MSSEG2016 train' (3021 slices): $0.630 > 0.564$ average Dice score. Alternatively, for homogeneous datasets, 3D-MR-MS (10517 slices) outranks ISBI2015 (2843 slices): $0.584 > 0.497$ average Dice score.

Finally, note that while the model trained on the large 'MSSEG2016 test' dataset is the top performer with an average Dice score of 0.630, the model trained on ISBI2015 displays the lowest performance (0.497 average Dice score) likely because it is both small and homogeneous (the difference is statistically significant at p=0.02714).

### 5.3 Increasing dataset size improves model performance

Next, we tested whether *explicitly* increasing the dataset size influences model performance. To achieve this, we considered combinations of datasets and compared the performance of the combined dataset with the performance of the individual datasets. For example, we compared the performance of a model trained on the datasets $A + B + C$ with the performances of models trained individually on A, B and C respectively (all of these models being tested on the same dataset D - compare column 'Dice score' with column 'Individual Dice scores' in Table 4).

Table 4 shows that *increasing dataset size by combining datasets produces slightly higher Dice scores* in all cases. For example:

$$\text{Dice(MSSEG2016-train+3D-MR-MS+ISBI2015, MSSEG2016-test)} = 0.631$$
$$> \quad \text{Dice(MSSEG2016-train, MSSEG2016-test)} = 0.602$$
$$\text{Dice(3D-MR-MS, MSSEG2016-test)} = 0.544$$
$$\text{Dice(ISBI2015, MSSEG2016-test)} = 0.516$$

where Dice(train,test) is the Dice score obtained by a model trained on 'train' and tested on 'test'.

As score improvements are expected to be only sublinear with dataset size increase, achieving significantly higher scores would probably require significantly larger datasets than the ones already available. Indeed, the gap between other vision datasets such as ImageNet with millions of examples and the MS datasets with only tens of thousands of slices is still too large and needs to be addressed somehow by more MS data, which can only be collected with a similar community effort in the future.

Also note that our model trained on 'MSSEG2016 train' combined with 3D-MR-MS and ISBI2015 surpasses the winner of the MICCAI 2016 competition (with a Dice score of 0.631, versus 0.5914 obtained by the winner) [7].

Given that the 'MSSEG2016 test' dataset is comprised of scans obtained in 4 distinct centers with 4 different scanners (one of which, namely center03, is completely distinct from the scanners used in the training dataset, even made by a different company, General Electric), we checked the Dice scores obtained by the above combined model averaged on the individual centers. We noticed that the performance of the model is largely consistent across the all different centers (0.664 for center01, 0.624 for center03, 0.627 for center07 and 0.608 for center08) and that the performance on the unseen center (center03) is close to the average performance of the model on the entire dataset, thereby confirming the generalizability of the model.

Table 4: Performance of models trained on combinations of datasets.

| Trained on | Tested on | Dice score | Individual Dice scores | IoU score |
|---|---|---|---|---|
| MSSEG-2016 train + 3D-MR-MS | MSSEG-2016 test | 0.624 | (0.602, 0.544) | 0.478 |
| MSSEG-2016 train + 3D-MR-MS | ISBI-2015 | 0.611 | (0.569, 0.603) | 0.450 |
| ISBI-2015 + MSSEG-2016 train | MSSEG-2016 test | 0.609 | (0.516, 0.602) | 0.463 |
| ISBI-2015 + MSSEG-2016 train | 3D-MR-MS | 0.533 | (0.46, 0.523) | 0.390 |
| ISBI-2015 + 3D-MR-MS | MSSEG-2016 test | 0.581 | (0.516, 0.544) | 0.438 |
| ISBI-2015 + 3D-MR-MS | MSSEG-2016 train | 0.639 | (0.515, 0.606) | 0.484 |
| MSSEG-2016 train + 3D-MR-MS + ISBI-2015 | MSSEG-2016 test | 0.631 | (0.602, 0.544, 0.516) | 0.487 |

### 5.4 Quality of annotations

The highest Dice score obtained in this paper is 0.698. Even the best state-of-the-art MS lesion segmentation models achieve average Dice scores around 0.67 [6, 12]. These still far from perfect scores could be due to the shortage of MS data, but also to the reduced quality of the annotations.

Surprisingly, we observed that expert annotators display only far from perfect agreement. Both ISBI2015 and MSSEG2016 datasets were annotated by multiple experts (2 in the case of ISBI2015 and 7 for MSSEG2016). While the MSSEG2016 dataset contains a consensus annotation computed with the Logarithmic Opinion Pool Based Simultaneous Truth And Performance Level Estimation (LOP STAPLE) algorithm [20], ISBI2015 has no such consensus annotation. In the case of MSSEG2016, the average Dice score between expert annotations and the consensus $mean_{scan} \, mean_i \, Dice(Expert_i, Consensus) = 0.72$ is significantly larger than the average Dice score between expert annotations $mean_{scan} \, mean_{i<j} \, Dice(Expert_i, Expert_j) = 0.61$, showing that the consensus annotation is of higher quality than the individual annotations. On the other hand, the Dice score between the 2 raters from ISBI2015 is $mean_{scan} \, Dice(Expert_1, Expert_2) = 0.73$.

To investigate to what extent annotation quality affects model performance, we trained models on ISBI2015 with annotations from expert 1, expert 2 and respectively the combined annotations of the two experts (their union). Table 5 shows that this simple consensus annotation leads to only marginally better results than the two individual expert annotations.[11] This rather small difference could be due to the very small sample size, but also to the difficulty of obtaining a significantly better consensus annotation from just two lower quality annotations. This reinforces once again the importance of obtaining high-quality annotations, which however may incur significant costs related to employing several independent annotators.

Therefore, besides dataset size, expert annotation quality also seems to be a major bottleneck towards significantly better MS lesion segmentation models. For example, it seems difficult to aim for Dice scores above 0.8, if annotation quality is not improved beyond the current 0.73.

Table 5: Results of models trained on ISBI2015 with first rater segmentation, second rater segmentation and their consensus (union)

| Trained on | Tested on | Dice score | IoU score |
|---|---|---|---|
| ISBI-2015 expert 1 | MSSEG-2016 train | 0.481 | 0.327 |
| ISBI-2015 expert 1 | MSSEG-2016 test | 0.511 | 0.359 |
| ISBI-2015 expert 1 | 3D-MR-MS | 0.394 | 0.257 |
| ISBI-2015 expert 2 | MSSEG-2016 train | 0.498 | 0.339 |
| ISBI-2015 expert 2 | MSSEG-2016 test | 0.506 | 0.354 |
| ISBI-2015 expert 2 | 3D-MR-MS | 0.423 | 0.279 |
| ISBI-2015 consensus | MSSEG-2016 train | 0.515 | 0.352 |
| ISBI-2015 consensus | MSSEG-2016 test | 0.516 | 0.362 |
| ISBI-2015 consensus | 3D-MR-MS | 0.460 | 0.273 |

### 5.5 Ablation studies

To demonstrate the contributions of the key components of our approach, we performed two ablation studies.

**The role of quantile normalization:** Table 6 compares the results of the models from Tables 2 and 4, obtained on quantile normalized datasets, with corresponding models trained and tested on the same datasets but which were linearly normalized to the [0,1] range instead of quantile normalized. A Wilcoxon paired statistical test revealed that quantile normalization significantly outperforms linear normalization (p=0.00003624).

**UNet++ versus UNet:** Table 7 compares models trained using the UNet++ architecture with models trained employing the UNet baseline. We observed that UNet++ performs better (statistically significant by a Wilcoxon paired test with p=0.00029).

---

[11]The Dice score average for the consensus model, 0.497, marginally exceeds the corresponding averages for the two raters 0.462, 0.476. However, a repeated measures ANOVA test on the Dice scores from Table 5 does not lead to the rejection of the null hypothesis of no difference between training models, p=0.1137.

Table 6: Ablation study for comparing quantile normalization with linear normalization (Dice scores)

| Trained on | Tested on | Quantile normalization | Linear normalization |
|---|---|---|---|
| MSSEG-2016 train | MSSEG-2016 test | 0.602 | 0.574 |
| MSSEG-2016 train | 3D-MR-MS | 0.523 | 0.510 |
| MSSEG-2016 train | ISBI-2015 | 0.569 | 0.535 |
| 3D-MR-MS | MSSEG-2016 test | 0.544 | 0.495 |
| 3D-MR-MS | MSSEG-2016 train | 0.606 | 0.549 |
| 3D-MR-MS | ISBI-2015 | 0.603 | 0.568 |
| MSSEG-2016 test | MSSEG-2016 train | 0.698 | 0.662 |
| MSSEG-2016 test | 3D-MR-MS | 0.569 | 0.502 |
| MSSEG-2016 test | ISBI-2015 | 0.623 | 0.553 |
| ISBI-2015 | MSSEG-2016 train | 0.515 | 0.475 |
| ISBI-2015 | MSSEG-2016 test | 0.516 | 0.460 |
| ISBI-2015 | 3D-MR-MS | 0.460 | 0.418 |
| MSSEG-2016 train + 3D-MR-MS | MSSEG-2016 test | 0.624 | 0.650 |
| MSSEG-2016 train + 3D-MR-MS | ISBI-2015 | 0.611 | 0.603 |
| ISBI-2015 + MSSEG-2016 train | MSSEG-2016 test | 0.609 | 0.607 |
| ISBI-2015 + MSSEG-2016 train | 3D-MR-MS | 0.533 | 0.459 |
| ISBI-2015 + 3D-MR-MS | MSSEG-2016 test | 0.581 | 0.558 |
| ISBI-2015 + 3D-MR-MS | MSSEG-2016 train | 0.639 | 0.613 |
| MSSEG-2016 train + 3D-MR-MS + ISBI-2015 | MSSEG-2016 test | 0.631 | 0.631 |

Table 7: Ablation study for comparing different model architectures, UNet++ vs. UNet (Dice scores)

| Trained on | Tested on | UNet++ | UNet |
|---|---|---|---|
| MSSEG-2016 train | MSSEG-2016 test | 0.602 | 0.596 |
| MSSEG-2016 train | 3D-MR-MS | 0.523 | 0.518 |
| MSSEG-2016 train | ISBI-2015 | 0.569 | 0.541 |
| 3D-MR-MS | MSSEG-2016 test | 0.544 | 0.518 |
| 3D-MR-MS | MSSEG-2016 train | 0.606 | 0.576 |
| 3D-MR-MS | ISBI-2015 | 0.603 | 0.569 |
| MSSEG-2016 test | MSSEG-2016 train | 0.698 | 0.676 |
| MSSEG-2016 test | 3D-MR-MS | 0.569 | 0.529 |
| MSSEG-2016 test | ISBI-2015 | 0.623 | 0.608 |
| ISBI-2015 | MSSEG-2016 train | 0.515 | 0.492 |
| ISBI-2015 | MSSEG-2016 test | 0.516 | 0.494 |
| ISBI-2015 | 3D-MR-MS | 0.460 | 0.390 |
| MSSEG-2016 train + 3D-MR-MS | MSSEG-2016 test | 0.624 | 0.622 |
| MSSEG-2016 train + 3D-MR-MS | ISBI-2015 | 0.611 | 0.594 |
| ISBI-2015 + MSSEG-2016 train | MSSEG-2016 test | 0.609 | 0.592 |
| ISBI-2015 + MSSEG-2016 train | 3D-MR-MS | 0.533 | 0.540 |
| ISBI-2015 + 3D-MR-MS | MSSEG-2016 test | 0.581 | 0.569 |
| ISBI-2015 + 3D-MR-MS | MSSEG-2016 train | 0.639 | 0.626 |
| MSSEG-2016 train + 3D-MR-MS + ISBI-2015 | MSSEG-2016 test | 0.631 | 0.624 |

## 6 Discussion and future work

Complementary to previous MS lesion segmentation challenges which focused on optimizing the performance on a single evaluation dataset, this study concentrates on developing models that perform well across a wide range of evaluation datasets (are 'generalizable'), a scenario closer to real life clinical usage, where much wider ranges of scanners, settings and patient cohorts are expected. We started by collecting all high-quality publicly available MS lesion segmentation datasets and systematically trained a state-of-the-art UNet++ architecture on each of these datasets. The resulting models showed similar performance across the remaining datasets, with larger and more heterogeneous datasets leading to better models. Next, we showed that *explicitly* increasing the dataset size by combining datasets also leads to improved performance. In particular, combining the 'MSSEG2016 train', ISBI2015 and 3D-MR-MS datasets produced a model with a Dice score of 0.631, surpassing the winner of the MICCAI 2016 competition using a simple method on a single MRI modality (FLAIR). Besides dataset size, the quality of expert annotations is also essential for model performance. Surprisingly, we observed that expert annotators display only far from perfect agreement in their lesion annotations (under 0.73).

Developing models that generalize out-of-distribution (OOD) is a crucial aspect in numerous applications [21], not just MS lesion segmentation. However, there seems to be no one-size-fits-all solution for OOD generalization that we could directly apply in our setting. Different techniques might be necessary depending on the specific characteristics of the OOD data and the task. In this work we discovered that a simple normalization method, quantile normalization, originally introduced in the context of genomic data, is an effective method for mitigating the distribution shifts due to different scanners and scanning protocols used in diverse MRI datasets. Simpler linear normalization methods do not perform as well in the case of MS lesions which involve hyperintensities from the tails of the distributions. To reduce dataset variability and enhance model robustness, we also standardize the scan orientations to work with axial slices. Therefore, two simple techniques, quantile normalization and working with standardized axial slices seem sufficient for ensuring robust OOD performance. However, other methods exist (e.g. [11, 12]) and could potentially be combined with our methods in future work.

The performance of state-of-the-art MS lesion segmentation models is still suboptimal for clinical applications [22]. Moreover, besides high performance on a specific clinical setting, an essential requirement for the wide adoption of a model is its robustness in widely different clinical settings (involving different scanners and scanner settings). Our work represents an important step toward this essential desideratum. Moreover, while previous lesion detection algorithms have demonstrated strong performance using internal datasets that are an order of magnitude larger and incorporating at least two MRI modalities (e.g., FLAIR and T1w) [12], we demonstrate that model generalizability can be achieved with a simpler method and just a single MRI modality (FLAIR) even with an order of magnitude less training data.

In conclusion, we expect deep learning based MS lesion segmentation models to become really useful in clinical settings only with more data, as well as higher quality annotations, which can only be obtained by a concerted community effort.

## Data and code availability

The datasets used in this study are available from the original websites

ISBI2015: `https://smart-stats-tools.org/lesion-challenge-2015`

MSSEG2016: `https://shanoir.irisa.fr/shanoir-ng/challenge-request`

3D-MR-MS: `https://lit.fe.uni-lj.si/en/research/resources/3D-MR-MS/`

Code and models: `https://github.com/marypopa/MS-LesionSegmentation-Xdataset`

## Author Contributions

**Liviu Badea**: Conceptualization, Methodology, Validation, Software, Investigation, Supervision, Writing - Review & Editing. **Maria Popa**: Software, Conceptualization, Methodology, Validation, Investigation, Writing-Original Draft.

## Acknowledgment

This work was partially supported by the project PN-2338-0501, financed by Ministry of Research, Innovation and Digitalization. We are deeply grateful to Dr. Vișa Gabriela (Neurology Department of the Pediatric Clinical Hospital Sibiu) for clinical insights.